\documentclass[aps,prl,twocolumn,showpacs,amsmath,preprintnumbers,amssymb,groupedaddress,superscriptaddress]{revtex4}
\usepackage{graphicx}
\usepackage{dcolumn}
\usepackage{bm}
\usepackage{multirow}
\usepackage{hhline}
\usepackage{bbm}

\begin{document}

\title{The roles of deformation and orientation in heavy-ion collisions induced by light deformed nuclei at intermediate energy}

\author{ X.G. Cao }
\affiliation{Shanghai Institute of Applied Physics, Chinese Academy
of Sciences, Shanghai 201800, China} \affiliation{Graduate School of
the Chinese Academy of Sciences, Beijing 100049, China}

\author{ G.Q. Zhang }
\affiliation{Shanghai Institute of Applied Physics, Chinese Academy
of Sciences, Shanghai 201800, China} \affiliation{Graduate School of
the Chinese Academy of Sciences, Beijing 100049, China}

\author{X.Z. Cai\footnote{E-mail address: caixz@sinap.ac.cn}}
\affiliation{Shanghai Institute of Applied Physics, Chinese
Academy of Sciences, Shanghai 201800, China}

\author{Y.G. Ma}
\affiliation{Shanghai Institute of Applied Physics, Chinese
Academy of Sciences, Shanghai 201800, China}

\author{W. Guo }
\affiliation{Shanghai Institute of Applied Physics, Chinese Academy
of Sciences, Shanghai 201800, China}

\author{J.G. Chen }
\affiliation{Shanghai Institute of Applied Physics, Chinese
Academy of Sciences, Shanghai 201800, China}

\author{W.D. Tian }
\affiliation{Shanghai Institute of Applied Physics, Chinese Academy
of Sciences, Shanghai 201800, China}

\author{D.Q. Fang }
\affiliation{Shanghai Institute of Applied Physics, Chinese Academy
of Sciences, Shanghai 201800, China}

\author{H.W. Wang }
\affiliation{Shanghai Institute of Applied Physics, Chinese
Academy of Sciences, Shanghai 201800, China}

\date{\today}

\begin{abstract}

The reaction dynamics of axisymmetric deformed $^{24}$Mg + $^{24}$Mg
collisions have been investigated systematically by an
isospin-dependent quantum molecular dynamics (IDQMD) model. It is
found that different deformations and orientations result in
apparently different properties of reaction dynamics. We revealed
that some observables such as nuclear stopping power ($R$),
multiplicity of fragments, and elliptic flow are very sensitive to
the initial deformations and orientations. There exists an
eccentricity scaling of elliptic flow in central body-body
collisions with different deformations. In addition, the tip-tip and
body-body configurations turn out to be two extreme cases in central
reaction dynamical process.

\end{abstract}


\maketitle

Aligned experiments investigating how deformed $^{165}$Ho target
affects the total neutron reaction cross section from 2 to 125 MeV
\cite{MH-PRL-20} and scattering of $\alpha$ particles with 15
\begin{math}\leq
\end{math} $E_\alpha$ \begin{math}\leq
\end{math} 23 MeV \cite{PDR-PRL-29} were carried out about forty years ago.
The similar case occurs in nanoscale physics that the initial shape
of hot droplets also has significant effects on fragmentation
process in the molecular dynamics (MD) framework \cite{KN-PRE-72}.
It is expected that deformed nuclei induced heavy-ion collisions
(HICs) can result in obviously different properties of dynamical
processes and final state observables compared with spherical cases.
There are some reports about deformed U + U collisions at
relativistic and ultrarelativistic energies and it is suggested that
deformed U + U collisions are more likely to create quark-gluon
plasma (QGP) and may resolve many outstanding problems
\cite{DGS-PRC-62,LBA-PRC-61,SEV-PRC-61,HU-PRL-94,KA-PRC-72,NC-PRC-76,LXF-PRC-76,MH-PLB-679}.
The deformation effects on reaction cross section \cite{CJA-PRC-59},
elliptic flow \cite{FP-PAN-71} and heavy-ion fusion
\cite{SRG-PRL-41,DVY-PRC-76} was also discussed recently. On the
other hand, polarized target and beam have been widely applied
related with spin effects in HICs \cite{FD-PS-214} especially for
the total and differential reaction cross section measurement of
aligned deformed beams such as $^{7}$Li \cite{MK-PRL-46} and
$^{23}$Na \cite{CNM-PRC-47}.

The spin polarized beams have been greatly promoted by
projectile-fragmentation reactions recently \cite{AK-PPNP-46}, which
brings large angular momentum into fragment spin. Not only the
fragmentation process itself produces spin polarized fragments but
also the produced spin orientated beam of deformed nuclei can
provide valuable information on shape effects during collisions
\cite{EW-2003}. Therefore, it is very necessary to consider the
degree of freedom of initial deformation since so many radioactive
nuclei far from $\beta$-stability line may have large deformation.
However, the knowledge about collisions induced by deformed nuclei
is very poor especially at intermediate energy.

Due to the distinct differences in overlap region of deformed nuclei
collisions, collisions of aligned deformed nuclei may give a clearer
and deeper insight into the reaction mechanism such as the process
of multifragmentation and the development of collective flow. The
different orientational collisions also have the advantage in fixing
the uncertain behavior of density dependent symmetry energy, which
is an elementary open problem related not only to many problems in
nuclear physics but also to a number of important issues in nuclear
astrophysics \cite{LBA-PR-464}. Besides the advantage in studying
reaction mechanism and dynamics, highly deformed nuclei induced
reactions may also inspire exotic nuclei research such as halo
\cite{FDQ-PRC-76} and cluster phenomena \cite{FM-RPP-70}.

In this paper, $^{24}$Mg + $^{24}$Mg collision system is used to
investigate the initial deformation and orientation effects by a
microscopic transport model: the IDQMD model \cite{MYG-PRC-73},
which was developed from the quantum molecular dynamics (QMD) model
\cite{AJ-PR-202}. The main advantage of the QMD model is that it can
explicitly treat the many body state of collision system. So it
contains correlation effects to all orders and can treat the
fragmentation and fluctuation of HICs well.

\begin{figure}
\centering \setlength{\belowcaptionskip}{-0.3cm}
\includegraphics[scale=0.5]{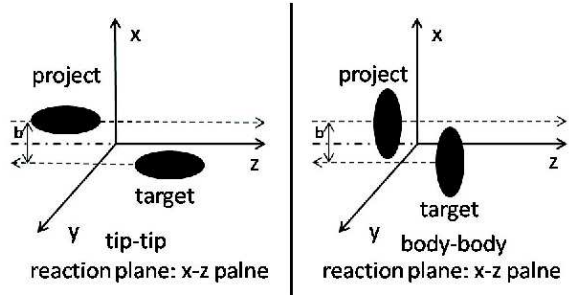}
\vspace{-0.1truein} \caption{\footnotesize Schematic representation
of tip-tip and body-body collisions. In the coordinate system,
$z$-direction is defined as the incident direction and the impact
parameter ($b$) is labeled as $x$-axis. Only $b$ to $^{24}$Mg's long
axis in body-body collisions is considered.}\label{schematic}
\end{figure}

In this calculations, soft and hard nuclear equation of state (EOS)
with the incompressibility of $K$ = 200 and 380 MeV, respectively,
are used for comparison. Here the strength of symmetry potential
$C_{sym}$ = 32 MeV \cite{AJ-PR-202} and experimental parameterized
nucleon-nucleon cross section which is energy and isospin dependent
are used. $^{24}$Mg is approximately treated as a sharp-cutoff
ellipsoid with large quadrupole deformation parameter: $\beta_{2}$ =
0.416 \cite{LGA-ADNDT-71}. For comparison, systematical calculations
for tip-tip (body-body) collisions of $^{24}$Mg + $^{24}$Mg with
$\beta_{2}$ = 0, 0.05, 0.1, 0.2 (all four cases with the same root
mean-square radius) at different energies and impact parameters are
carried out . The schematic plot of tip-tip and body-body collisions
is illustrated by Fig. \ref{schematic}.

\begin{figure}
\centering \setlength{\belowcaptionskip}{-0.3cm}
\includegraphics[width=8.6cm]{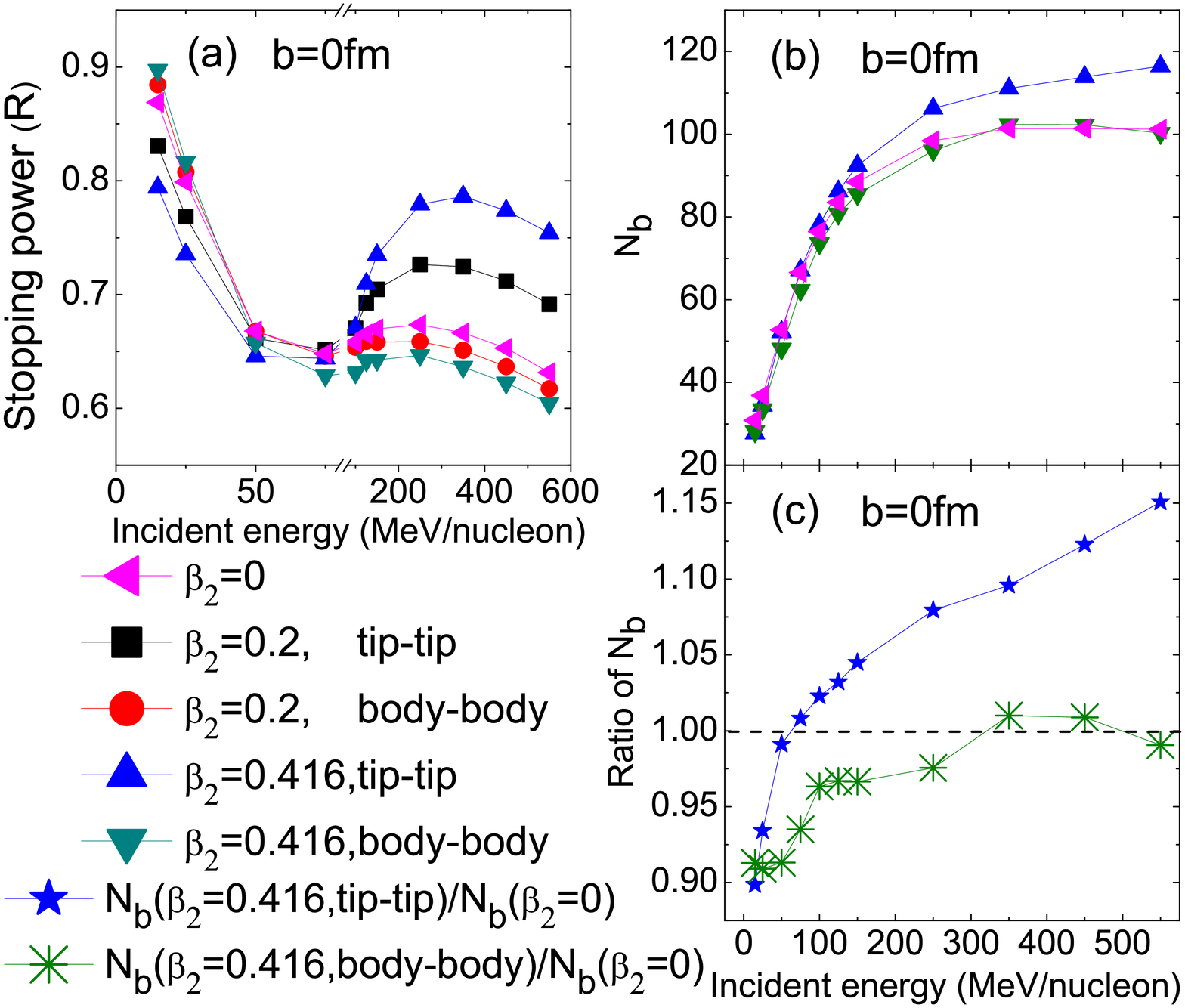}
\vspace{-0.1truein} \caption{\footnotesize (Color online) (a): $R$
and (b): total binary collision number ($N_b$) as a function of
incident energy at freeze-out time. (c): The ratio of non-spherical
$N_b$ to spherical one.}\label{stopping_ncoll}
\end{figure}

Firstly, we discuss the nuclear stopping power (\begin{math} R =
\frac{2}{\pi}\sum_{i}^A\mid{P_{i\perp}}\mid/\sum_{i}^A\mid{P_{i\parallel}}\mid
\end{math}, where A refers to the sum of projectile mass number and target mass
number, \begin{math} P_{i\perp} = (P_{i x}^2 + P_{i y}^2)^{1/2},
\end{math} \begin{math} P_{i\parallel} = P_{i z}
\end{math} in the c.m. reference system \cite{SH-PRC-27}) of different orientational collisions.
$R$ can be used to describe the momentum dissipation and the degree
of thermalization. Fig. \ref{stopping_ncoll} (a) shows that central
body-body collisions lead to larger $R$ than central tip-tip
collisions below 50MeV/nucleon while the situation reverses when
incident energies exceed 75MeV/nucleon. The more prolate $^{24}$Mg
is, more obvious differences appear. The spherical case lies between
tip-tip and body-body collisions at all calculated energies. The
larger $R$ of tip-tip collisions at higher energy is in agreement
with the result at 0.52GeV/nucleon by ART model \cite{LXF-PRC-76}.
However, the inversion of $R$ between tip-tip and body-body
collisions is first observed. It reflects the different roles of the
initial space configurations vs. energies.
\begin{figure}
\centering \setlength{\belowcaptionskip}{-0.3cm}
\includegraphics[width=8.6cm]{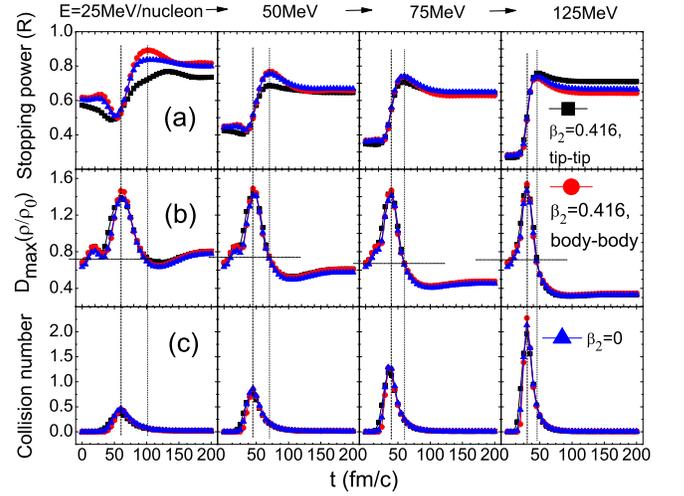}
\vspace{-0.1truein} \caption{\footnotesize (Color online) (a), (b)
and (c) represent the time evolutions of $R$, maximal density
$D_{max}$ and $N_b$ in central collisions($b = 0fm$), respectively.
The long-dashed, short-dashed and dash-dotted lines are drawn to
mark the characteristic time of the collisions. The time structure
of the $N_b$ is synchronous with the density evolution. Time
evolution of $R$ show tip-tip, body-body and sphere-sphere
collisions experience different touching, compressing and expanding
processes from $t$ = 0 to freeze-out stage.}\label{stopping_t}
\end{figure}

It is known that the reaction mechanism at intermediate energy is
dominated by mean field, binary collisions and Pauli blocking. Since
the IDQMD model can treat the three components explicitly, it is
very convenient to find out the factors which dominate the stopping
power at different energies. As represented in Fig.
\ref{stopping_ncoll} (b) and (c), tip-tip collision numbers are
higher than body-body ones at all considered energies. It means that
binary collision cannot be responsible for the inversion of $R$,
while the mean field must play a very important role. Fig.
\ref{stopping_t} shows how mean field and binary collision take
effect in dynamical process. The peak of density corresponds to the
most intensive stopping process but the $R$ has not reached maximum.
The departure between projectile-like and target-like continues
contributing the nuclear stopping power. Through the different
stopping behaviors of tip-tip and body-body collisions vs. energies,
the time evolutions of $R$ show that when reaction proceeds more
quickly, the larger stopping power can be achieved. So the stopping
power can be regarded as a measurement of time scale of dynamical
process as well as an observable of momentum dissipation.

Due to the larger projectile-target overlap region, body-body
configurations build up stronger mean field, which lead to more
violent one-body scattering. However, transparency effect of the
nuclear medium becomes more and more important when incident energy
rises. The tip-tip configurations are less transparent, which leads
to larger two-body collisions and stopping power. Therefore, the
underlying mechanics of the inversion of $R$ between tip-tip and
body-body collisions is that body-body configurations build up
stronger mean field at lower energies, where one-body scattering is
predominant. Whereas two-body collisions become more important in
tip-tip configurations at higher energies.

\begin{figure}
\centering
\setlength{\belowcaptionskip}{-0.3cm}
\includegraphics[width=8.6cm]{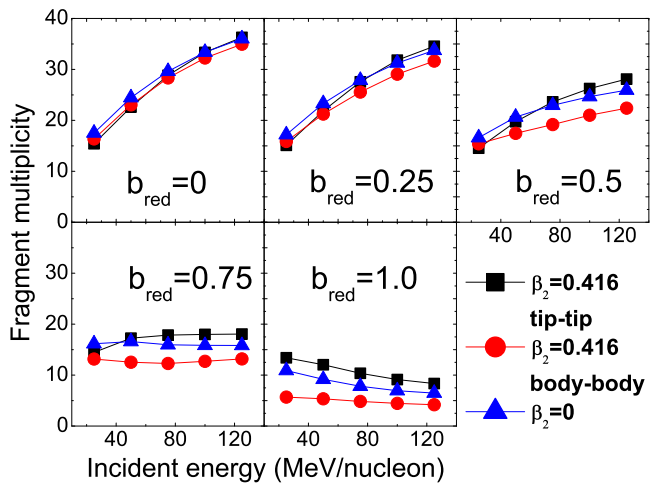}
\vspace{-0.1truein} \caption{\footnotesize (Color online) Energy
dependence of fragment multiplicity of tip-tip, body-body and
sphere-sphere collisions at different reduced impact parameters
($b_{red}$ = $b$/$b_{max}$, where $b_{max}$ refers to the maximal
impact parameter for different cases). }\label{mult}
\end{figure}

\begin{figure}
\centering \setlength{\belowcaptionskip}{-0.3cm}
\includegraphics[width=8.6cm]{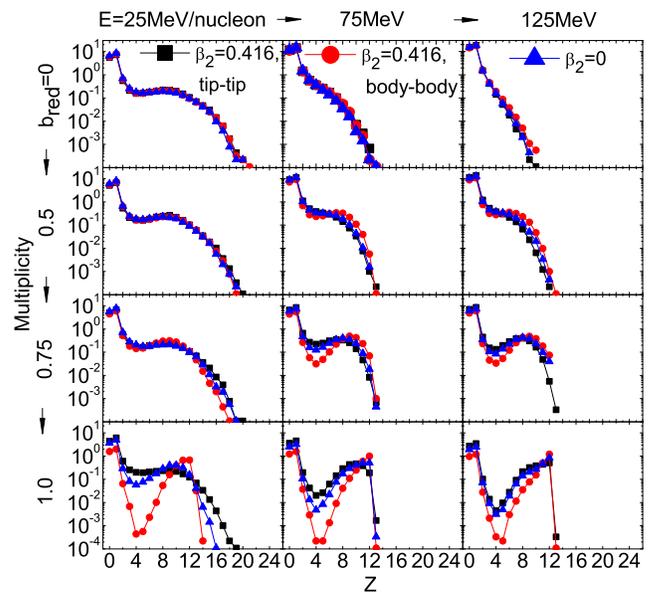}
\vspace{-0.1truein} \caption{\footnotesize (Color online) Charge
distributions of tip-tip, body-body and sphere-sphere collisions at
different $b_{red}$ and incident energies. }\label{mult_dis}
\end{figure}

Since the IDQMD model can treat fragmentation of hot nuclei
\cite{MYG-PRL-83,BA-RNC-23} well. It is appropriate to investigate
the fragmentation observables. As shown in Fig. \ref{mult}, the
fragment multiplicity has strong correlation with stopping power.
Body-body collisions have minimal multiplicity at all impact
parameters at higher energies while tip-tip collisions have the
maximal one. So this behavior is consistent with that of stopping
power at higher energies. It can also be seen from charge
distributions in Fig. \ref{mult_dis} that the tip-tip and body-body
collisions are two extreme cases and the sphere-sphere collisions
lie between them. Therefore, the fragment observables also confirm
the similar picture indicated by $R$.

The Body-body collisions with $b$ = 0 $fm$ will produce large
collective motions due to the different initial geometry from
spherical nuclei. Anisotropic flow method has been developed to
measure the anisotropy of particle momentum space which related to
the EOS and nuclear reaction dynamics
\cite{DP-S-298,FPC-AA-PLB-612,YTZ-PLB-638}. The azimuthal
distribution of fragments can be expressed by Fourier expansion
\cite{VS-ZPC-70}
\begin{math}
\frac{dN}{d\phi} \propto {1 + 2\sum_{n=1}^\infty v_n cos(n\phi) },
\end{math}
where $\phi$ is azimuthal angle between the transverse momentum of
the particle and the reaction plane. The coefficient $v_{n}$ is
defined as anisotropic flow parameter, among which $v_{2}$ denotes
elliptic flow. It can be calculated in terms of single-particle
averages:
\begin{math}
v_2 = \langle cos(2\phi) \rangle = \langle
\frac{p^2_x-p^2_y}{p_x^2+p_y^2} \rangle.
\end{math}
Nucleon's $v_2$ induced by deformed U + U collisions has been
studied by ART model \cite{LBA-PRC-61,LXF-PRC-76} and optical
Glauber model \cite{FP-PAN-71} at relativistic energies recently. It
seems that the most central body-body collisions give rise to
largest $v_2$ because of the strongest shadowing effect in the
reaction plane \cite{LBA-PRC-61}. Thus $v_{2}$ of central body-body
collisions are most appropriate for investigating the EOS. However,
$v_2$ developed from deformed nuclei collisions is unknown at
intermediate energy and it's interesting to study their deformation
and orientation effects.

\begin{figure}
\centering \setlength{\belowcaptionskip}{-0.3cm}
\includegraphics[width=8.6cm]{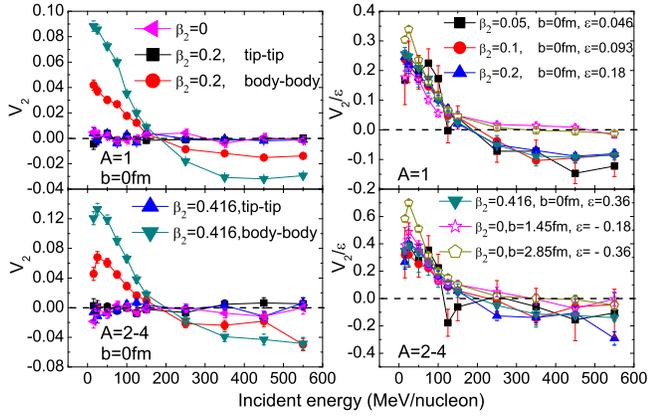}
\vspace{-0.1truein} \caption{\footnotesize (Color online) $v_{2}$
excitation function of light fragments at mid-rapidity ($-0.5<Y<
0.5$) of deformed and spherical collisions. The dashed lines are
drawn to guide the eye. Left: $v_{2}$ in central collisions with $b$
= 0 $fm$; Right: scaled $v_{2}$ with eccentricity $\epsilon$ in
central body-body collisions and non-central spherical $^{24}$Mg
collisions. The spherical $^{24}$Mg collisions with $b$ = 1.45 $fm$
and 2.85 $fm$ have the same absolute value of $\epsilon$ as the
deformed central $^{24}$Mg collisions with $\beta_{2}$ = 0.2 and
0.416 , respectively.}\label{V2_soft_eos}
\end{figure}

$v_{2}$ of light fragments are shown in Fig. \ref{V2_soft_eos}, in
which the eccentricity ($\epsilon$) is calculated by maximal
geometry overlap region:
\begin{math}
\epsilon =
\sum\limits_{i}(x_{i}^{2}-y_{i}^{2})/\sum\limits_{i}(x_{i}^{2}+y_{i}^{2}).
\end{math}
Cental tip-tip and sphere-sphere collisions do not have obvious
$v_2$ because of the transverse symmetry of overlap region while the
$v_{2}$ of cental body-body collisions has non-zero value. The
negative sign of $v_{2}$ at higher energies is in agreement with
deformed U + U collisions by ART model \cite{LBA-PRC-61,LXF-PRC-76}.
The positive $v_{2}$ at lower energies and the alteration of sign
for $v_{2}$ are first observed in central body-body collisions. At
higher energies the violent two-body collisions in overlap region
build the anisotropy pressure and it prompts fragments emission from
in-plane preferential to out-of-plane preferential. The heavier
fragments have larger $v_{2}$, which is consistent with ref.
\cite{BD-ZP-355}. $v_{2}$ of central body-body collisions
($\beta_{2}$ = 0.05, 0.1, 0.2, 0.416) can be scaled together by
$\epsilon$ from low energies to high energies. While scaled by the
same $\epsilon$ amplitude as the deformed $^{24}$Mg collisions,
$v_{2}$ for mid-central spherical $^{24}$Mg collisions shows
different behaviors especially for higher energies. Therefore, the
scaling of $v_{2}$ indicates that the geometric shapes of
participants play an essential role in collective flow of central
body-body collisions.

\begin{figure}
\centering \setlength{\belowcaptionskip}{-0.3cm}
\includegraphics[width=8.6cm]{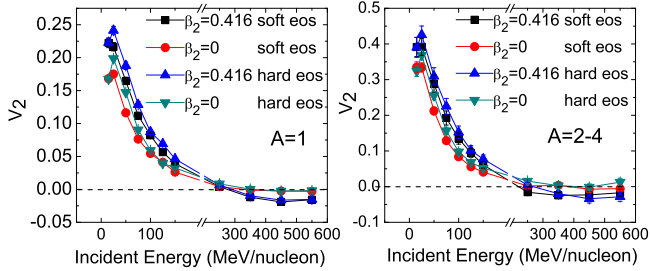}
\vspace{-0.1truein} \caption{\footnotesize (Color online) Average
$v_{2}$ excitation function of light fragments at mid-rapidity
($-0.5<Y< 0.5$) for deformed and spherical collisions with soft and
hard EOS. $v_{2}$ is averaged with $b$ from 0 to $b_{max}$ for
body-body and sphere-sphere collisions.}\label{V2_soft_hard_eos}
\end{figure}

The energy excitation function of $v_{2}$ at mid-central
sphere-sphere collisions varies from positive (in-plane,
rotational-like emission) to negative (out-of-plane, ``squeeze-out"
pattern) \cite{GHH-PLB-216,FPC-AA-PLB-612}. This energy point is
so-called transition energy, which is near 100MeV/nucleon
\cite{FPC-AA-PLB-612}. For spherical collision system, there exist
three competing components affecting the transition energy: (1)
rotation of the compound system, (2) expansion of the hot and
compressed participant zone, (3) shadowing of the colder spectator
region \cite{FPC-AA-PLB-612}. Only the expansion survives in central
spherical collisions\cite{SPJ-PRL-42,RW-NPA-612}, which merely
generates azimuthal symmetric flow. However, central body-body
collisions have bulk transverse asymmetry overlap region and there
is no rotation effect. Also the shadowing is different from
mid-central collisions of spherical nuclei. Therefore, it provides
an ideal tool to understand how the azimuthal pressure, expansion
and flow development from the almond-shape overlap, which are all
related with the extraction of the EOS. Average $v_{2}$ is shown in
Fig. \ref{V2_soft_hard_eos} with soft and hard EOS. Hard EOS
enhances $v_{2}$ for both spherical and deformed collisions.
Deformed configuration gives rise to larger $v_{2}$ than
sphere-sphere configuration for both soft and hard EOS.

In summary, deformed $^{24}$Mg + $^{24}$Mg collisions have been
studied systematically by IDQMD model. The inversion of $R$ vs.
energies between tip-tip and body-body collisions reflects the two
different configurations play different roles on reaction dynamics.
The fragment observables also show different behaviors for the two
configurations. The sphere-sphere collisions lie between tip-tip and
body-body collisions in nuclear stopping and fragmentation.
Moreover, the excitation functions of $v_2$ for different deformed
central body-body collisions can be scaled on a similar curve by
eccentricity. $v_2$ averaged by impact parameter (collision
configuration is represented by Fig. \ref{schematic}) in deformed
collisions is stronger than that of spherical collisions for both
soft and hard EOS. The large $v_2$ developed from cental body-body
collisions have advantages in studying the EOS and transition
energy. Tip-tip collisions can be used to study the liquid-gas phase
transition in finite nuclear systems due to the longer collision
time. In addition, deformed nuclei collisions may have some
implications on halo and cluster structure research. Therefore, the
merits of collisions of deformed nuclei can shed light on the
studies of both the nuclear structure and the reaction dynamics from
low energies to relativistic energies.

This work is partially supported by National Natural Science
Foundation of China under contract No.s 10775167, 10775168,
10979074, 10605036, 10875160, 10805067 and 10975174, Major State
Basic Research Development Program in China under contract No.
2007CB815004, the Shanghai Development Foundation for Science and
Technology under contract No. 09JC1416800 and the National Defence
Innovation Foundation of Chinese Academy of Science under grants No.
CXJJ-216.

\vskip -0.5cm



\begin{thebibliography}{37}
\expandafter\ifx\csname
natexlab\endcsname\relax\def\natexlab#1{#1}\fi
\expandafter\ifx\csname bibnamefont\endcsname\relax
  \def\bibnamefont#1{#1}\fi
\expandafter\ifx\csname bibfnamefont\endcsname\relax
  \def\bibfnamefont#1{#1}\fi
\expandafter\ifx\csname citenamefont\endcsname\relax
  \def\citenamefont#1{#1}\fi
\expandafter\ifx\csname url\endcsname\relax
  \def\url#1{\texttt{#1}}\fi
\expandafter\ifx\csname urlprefix\endcsname\relax\def\urlprefix{URL
}\fi \providecommand{\bibinfo}[2]{#2}
\providecommand{\eprint}[2][]{\url{#2}}

\bibitem[{\citenamefont{Marshak et~al.}(1968)}]{MH-PRL-20}
\bibinfo{author}{\bibfnamefont{H.}~\bibnamefont{Marshak}} \bibnamefont{et~al.},
  \bibinfo{journal}{Phys. Rev. Lett.} \textbf{\bibinfo{volume}{20}},
  \bibinfo{pages}{554} (\bibinfo{year}{1968}).

\bibitem[{\citenamefont{Parks et~al.}(1972)}]{PDR-PRL-29}
\bibinfo{author}{\bibfnamefont{D.~R.} \bibnamefont{Parks}}
  \bibnamefont{et~al.}, \bibinfo{journal}{Phys. Rev. Lett.}
  \textbf{\bibinfo{volume}{29}}, \bibinfo{pages}{1264} (\bibinfo{year}{1972}).

\bibitem[{\citenamefont{Komatsu and Abe}(2005)}]{KN-PRE-72}
\bibinfo{author}{\bibfnamefont{N.}~\bibnamefont{Komatsu}} \bibnamefont{and}
  \bibinfo{author}{\bibfnamefont{T.}~\bibnamefont{Abe}},
  \bibinfo{journal}{Phys. Rev. E} \textbf{\bibinfo{volume}{72}},
  \bibinfo{pages}{021601} (\bibinfo{year}{2005}).

\bibitem[{\citenamefont{Das~Gupta and Gale}(2000)}]{DGS-PRC-62}
\bibinfo{author}{\bibfnamefont{S.}~\bibnamefont{Das~Gupta}} \bibnamefont{and}
  \bibinfo{author}{\bibfnamefont{C.}~\bibnamefont{Gale}},
  \bibinfo{journal}{Phys. Rev. C} \textbf{\bibinfo{volume}{62}},
  \bibinfo{pages}{031901(R)} (\bibinfo{year}{2000}).

\bibitem[{\citenamefont{Li}(2000)}]{LBA-PRC-61}
\bibinfo{author}{\bibfnamefont{B.-A.} \bibnamefont{Li}},
  \bibinfo{journal}{Phys. Rev. C} \textbf{\bibinfo{volume}{61}},
  \bibinfo{pages}{021903(R)} (\bibinfo{year}{2000}).

\bibitem[{\citenamefont{Shuryak}(2000)}]{SEV-PRC-61}
\bibinfo{author}{\bibfnamefont{E.~V.} \bibnamefont{Shuryak}},
  \bibinfo{journal}{Phys. Rev. C} \textbf{\bibinfo{volume}{61}},
  \bibinfo{pages}{034905} (\bibinfo{year}{2000}).

\bibitem[{\citenamefont{Heinz and Kuhlman}(2005)}]{HU-PRL-94}
\bibinfo{author}{\bibfnamefont{U.}~\bibnamefont{Heinz}} \bibnamefont{and}
  \bibinfo{author}{\bibfnamefont{A.}~\bibnamefont{Kuhlman}},
  \bibinfo{journal}{Phys. Rev. Lett.} \textbf{\bibinfo{volume}{94}},
  \bibinfo{pages}{132301} (\bibinfo{year}{2005}).

\bibitem[{\citenamefont{Kuhlman and Heinz}(2005)}]{KA-PRC-72}
\bibinfo{author}{\bibfnamefont{A.}~\bibnamefont{Kuhlman}} \bibnamefont{and}
  \bibinfo{author}{\bibfnamefont{U.}~\bibnamefont{Heinz}},
  \bibinfo{journal}{Phys. Rev. C} \textbf{\bibinfo{volume}{72}},
  \bibinfo{pages}{037901} (\bibinfo{year}{2005}).

\bibitem[{\citenamefont{Nepali et~al.}(2007)\citenamefont{Nepali, Fai, and
  Keane}}]{NC-PRC-76}
\bibinfo{author}{\bibfnamefont{C.}~\bibnamefont{Nepali}},
  \bibinfo{author}{\bibfnamefont{G.}~\bibnamefont{Fai}}, \bibnamefont{and}
  \bibinfo{author}{\bibfnamefont{D.}~\bibnamefont{Keane}},
  \bibinfo{journal}{Phys. Rev. C} \textbf{\bibinfo{volume}{76}},
  \bibinfo{pages}{051902(R)} (\bibinfo{year}{2007}).

\bibitem[{\citenamefont{Luo et~al.}(2007)}]{LXF-PRC-76}
\bibinfo{author}{\bibfnamefont{X.-F.} \bibnamefont{Luo}} \bibnamefont{et~al.},
  \bibinfo{journal}{Phys. Rev. C} \textbf{\bibinfo{volume}{76}},
  \bibinfo{pages}{044902} (\bibinfo{year}{2007}).

\bibitem[{\citenamefont{Masui et~al.}(2009)}]{MH-PLB-679}
\bibinfo{author}{\bibfnamefont{H.}~\bibnamefont{Masui}} \bibnamefont{et~al.},
  \bibinfo{journal}{Phys. Lett. B} \textbf{\bibinfo{volume}{679}},
  \bibinfo{pages}{440} (\bibinfo{year}{2009}).

\bibitem[{\citenamefont{Christley and Tostevin}(1999)}]{CJA-PRC-59}
\bibinfo{author}{\bibfnamefont{J.~A.} \bibnamefont{Christley}}
  \bibnamefont{and} \bibinfo{author}{\bibfnamefont{J.~A.}
  \bibnamefont{Tostevin}}, \bibinfo{journal}{Phys. Rev. C}
  \textbf{\bibinfo{volume}{59}}, \bibinfo{pages}{2309} (\bibinfo{year}{1999}).

\bibitem[{\citenamefont{Filip}(2008)}]{FP-PAN-71}
\bibinfo{author}{\bibfnamefont{P.}~\bibnamefont{Filip}},
  \bibinfo{journal}{Phys. At. Nucl.} \textbf{\bibinfo{volume}{71}},
  \bibinfo{pages}{1609} (\bibinfo{year}{2008}).

\bibitem[{\citenamefont{Stokstad et~al.}(1978)}]{SRG-PRL-41}
\bibinfo{author}{\bibfnamefont{R.~G.} \bibnamefont{Stokstad}}
  \bibnamefont{et~al.}, \bibinfo{journal}{Phys. Rev. Lett.}
  \textbf{\bibinfo{volume}{41}}, \bibinfo{pages}{465} (\bibinfo{year}{1978}).

\bibitem[{\citenamefont{Denisov and Pilipenko}(2007)}]{DVY-PRC-76}
\bibinfo{author}{\bibfnamefont{V.~Y.} \bibnamefont{Denisov}} \bibnamefont{and}
  \bibinfo{author}{\bibfnamefont{N.~A.} \bibnamefont{Pilipenko}},
  \bibinfo{journal}{Phys. Rev. C} \textbf{\bibinfo{volume}{76}},
  \bibinfo{pages}{014602} (\bibinfo{year}{2007}).

\bibitem[{\citenamefont{Fick et~al.}(1992)}]{FD-PS-214}
\bibinfo{author}{\bibfnamefont{D.}~\bibnamefont{Fick}} \bibnamefont{et~al.},
  \bibinfo{journal}{Phys. Rep.} \textbf{\bibinfo{volume}{214}},
  \bibinfo{pages}{1} (\bibinfo{year}{1992}).

\bibitem[{\citenamefont{M\"{o}bius et~al.}(1981)}]{MK-PRL-46}
\bibinfo{author}{\bibfnamefont{K.~H.} \bibnamefont{M\"{o}bius}}
  \bibnamefont{et~al.}, \bibinfo{journal}{Phys. Rev. Lett.}
  \textbf{\bibinfo{volume}{46}}, \bibinfo{pages}{1064} (\bibinfo{year}{1981}).

\bibitem[{\citenamefont{Clarke et~al.}(1993)}]{CNM-PRC-47}
\bibinfo{author}{\bibfnamefont{N.~M.} \bibnamefont{Clarke}}
  \bibnamefont{et~al.}, \bibinfo{journal}{Phys. Rev. C}
  \textbf{\bibinfo{volume}{47}}, \bibinfo{pages}{660} (\bibinfo{year}{1993}).

\bibitem[{\citenamefont{Asahi}(2001)}]{AK-PPNP-46}
\bibinfo{author}{\bibfnamefont{K.}~\bibnamefont{Asahi}},
  \bibinfo{journal}{Prog. Part. Nucl. Phys.} \textbf{\bibinfo{volume}{46}},
  \bibinfo{pages}{321} (\bibinfo{year}{2001}).

\bibitem{EW-2003} {Exploratory Workshop on Polarized Radioactive Beams \\and Polarized Targets, 6-7 March 2003, Strasbourg,\\ Frances;
http://wwwires.in2p3.fr/ires/workshops/polar03/}.

\bibitem[{\citenamefont{Li et~al.}(2008)}]{LBA-PR-464}
\bibinfo{author}{\bibfnamefont{B.-A.} \bibnamefont{Li}} \bibnamefont{et~al.},
  \bibinfo{journal}{Phys. Rep.} \textbf{\bibinfo{volume}{464}},
  \bibinfo{pages}{113} (\bibinfo{year}{2008}).

\bibitem[{\citenamefont{Fang et~al.}(2007)}]{FDQ-PRC-76}
\bibinfo{author}{\bibfnamefont{D.~Q.} \bibnamefont{Fang}} \bibnamefont{et~al.},
  \bibinfo{journal}{Phys. Rev. C} \textbf{\bibinfo{volume}{76}},
  \bibinfo{pages}{031601(R)} (\bibinfo{year}{2007}).

\bibitem[{\citenamefont{Freer}(2007)}]{FM-RPP-70}
\bibinfo{author}{\bibfnamefont{M.}~\bibnamefont{Freer}}, \bibinfo{journal}{Rep.
  Prog. Phys.} \textbf{\bibinfo{volume}{70}}, \bibinfo{pages}{2149}
  (\bibinfo{year}{2007}).

\bibitem[{\citenamefont{Ma et~al.}(2006)}]{MYG-PRC-73}
\bibinfo{author}{\bibfnamefont{Y.~G.} \bibnamefont{Ma}} \bibnamefont{et~al.},
  \bibinfo{journal}{Phys. Rev. C} \textbf{\bibinfo{volume}{73}}
  \bibinfo{pages}{014604} (\bibinfo{year}{2006}).

\bibitem[{\citenamefont{Aichelin}(1991)}]{AJ-PR-202}
\bibinfo{author}{\bibfnamefont{J.}~\bibnamefont{Aichelin}},
  \bibinfo{journal}{Phys. Rep.} \textbf{\bibinfo{volume}{202}},
  \bibinfo{pages}{233} (\bibinfo{year}{1991}).

\bibitem[{\citenamefont{Lalazissis et~al.}(1999)}]{LGA-ADNDT-71}
\bibinfo{author}{\bibfnamefont{G.~A.} \bibnamefont{Lalazissis}}
  \bibnamefont{et~al.}, \bibinfo{journal}{At. Data Nucl. Data Tables}
  \textbf{\bibinfo{volume}{71}}, \bibinfo{pages}{1} (\bibinfo{year}{1999}).

\bibitem[{\citenamefont{Str\"{o}bele et~al.}(1983)}]{SH-PRC-27}
\bibinfo{author}{\bibfnamefont{H.}~\bibnamefont{Str\"{o}bele}}
  \bibnamefont{et~al.}, \bibinfo{journal}{Phys. Rev. C}
  \textbf{\bibinfo{volume}{27}}, \bibinfo{pages}{1349} (\bibinfo{year}{1983}).

\bibitem[{\citenamefont{Ma}(1999)}]{MYG-PRL-83}
\bibinfo{author}{\bibfnamefont{Y.~G.} \bibnamefont{Ma}},
  \bibinfo{journal}{Phys. Rev. Lett.} \textbf{\bibinfo{volume}{83}},
  \bibinfo{pages}{3617} (\bibinfo{year}{1999}).

\bibitem[{\citenamefont{Bonasera et~al.}(2000)}]{BA-RNC-23}
\bibinfo{author}{\bibfnamefont{A.}~\bibnamefont{Bonasera}}
  \bibnamefont{et~al.}, \bibinfo{journal}{Rivista del Nuovo Cimento}
  \textbf{\bibinfo{volume}{23}}, \bibinfo{pages}{1} (\bibinfo{year}{2000}).

\bibitem[{\citenamefont{Danielewicz et~al.}(2002)}]{DP-S-298}
\bibinfo{author}{\bibfnamefont{P.}~\bibnamefont{Danielewicz}}
  \bibnamefont{et~al.}, \bibinfo{journal}{Science}
  \textbf{\bibinfo{volume}{298}}, \bibinfo{pages}{1592} (\bibinfo{year}{2002}).

\bibitem[{\citenamefont{Andronic et~al.}(2005)}]{FPC-AA-PLB-612}
\bibinfo{author}{\bibfnamefont{A.}~\bibnamefont{Andronic}}
  \bibnamefont{et~al.}, \bibinfo{journal}{Phys. Lett. B}
  \textbf{\bibinfo{volume}{612}}, \bibinfo{pages}{173} (\bibinfo{year}{2005}).

\bibitem[{\citenamefont{Yan et~al.}(2006)}]{YTZ-PLB-638}
\bibinfo{author}{\bibfnamefont{T.~Z.} \bibnamefont{Yan}} \bibnamefont{et~al.},
  \bibinfo{journal}{Phys. Lett. B} \textbf{\bibinfo{volume}{638}},
  \bibinfo{pages}{50} (\bibinfo{year}{2006}).

\bibitem[{\citenamefont{Voloshin et~al.}(1996)}]{VS-ZPC-70}
\bibinfo{author}{\bibfnamefont{S.}~\bibnamefont{Voloshin}}
  \bibnamefont{et~al.}, \bibinfo{journal}{Z. Phys. C}
  \textbf{\bibinfo{volume}{70}}, \bibinfo{pages}{665} (\bibinfo{year}{1996}).

\bibitem[{\citenamefont{Brill et~al.}(1996)}]{BD-ZP-355}
\bibinfo{author}{\bibfnamefont{D.}~\bibnamefont{Brill}} \bibnamefont{et~al.},
  \bibinfo{journal}{Z. Phys. A} \textbf{\bibinfo{volume}{355}},
  \bibinfo{pages}{61} (\bibinfo{year}{1996}).

\bibitem[{\citenamefont{Gutbrod et~al.}(1989)}]{GHH-PLB-216}
\bibinfo{author}{\bibfnamefont{H.~H.} \bibnamefont{Gutbrod}}
  \bibnamefont{et~al.}, \bibinfo{journal}{Phys. Lett. B}
  \textbf{\bibinfo{volume}{216}}, \bibinfo{pages}{267} (\bibinfo{year}{1989}).

\bibitem[{\citenamefont{Siemens and Rasmussen}(1979)}]{SPJ-PRL-42}
\bibinfo{author}{\bibfnamefont{P.~J.} \bibnamefont{Siemens}} \bibnamefont{and}
  \bibinfo{author}{\bibfnamefont{J.~O.} \bibnamefont{Rasmussen}},
  \bibinfo{journal}{Phys. Rev. Lett.} \textbf{\bibinfo{volume}{42}},
  \bibinfo{pages}{880} (\bibinfo{year}{1979}).

\bibitem[{\citenamefont{Reisdorf et~al.}(1997)}]{RW-NPA-612}
\bibinfo{author}{\bibfnamefont{W.}~\bibnamefont{Reisdorf}}
  \bibnamefont{et~al.}, \bibinfo{journal}{Nucl. Phys. A}
  \textbf{\bibinfo{volume}{612}}, \bibinfo{pages}{493} (\bibinfo{year}{1997}).

\end{thebibliography}

\end{document}